\documentclass[twoside]{article}
\usepackage{fleqn,espcrc2}
\usepackage{graphicx}
\usepackage{amsmath}
\usepackage{amsfonts}
\usepackage{slashed}
\usepackage{cite}
\newcommand{\be}{\begin{equation}}
\newcommand{\ee}{\end{equation}}
\newcommand{\bdm}{\begin{displaymath}}
\newcommand{\edm}{\end{displaymath}}
\newcommand{\bea}{\begin{eqnarray}}
\newcommand{\eea}{\end{eqnarray}}
\newcommand{\ba}{\begin{array}}
\newcommand{\ea}{\end{array}}
\newcommand{\nn}{\nonumber}
\newcommand{\pref}[1]{(\ref{#1})}
\newcommand{\chpt}{$\chi$PT}

\title{Chiral perturbation theory at non-zero lattice spacing
}
\author{Oliver B\"ar\address{Graduate School of Pure and Applied Sciences, University of Tsukuba, Tsukuba 305-8571, Japan}
 } 
\begin{document}
\begin{abstract}
A review of chiral perturbation theory for lattice QCD at non-zero lattice spacing is given. 
\vspace{0.8pc}
\end{abstract}

\maketitle

\section{Introduction}
%
In spite of constant progress in computer technology, numerical lattice simulations with quark masses as light as realized in nature are out of reach. The smallest values for the ratio $M_{\pi}/M_{\rho}$ reported by various collaborations  during this conference  \cite{MpiOverMrhoLat2004} range from 0.35 to 0.66 for Wilson fermions with 2 flavors to 0.62 for those with 2+1 flavors. Simulations with staggered fermions reached a  value of 0.3, and 2 flavor domain-wall fermions were performed at a value of 0.53. 
All these numbers are still far away from the physical value 0.18. Consequently, numerical lattice simulations still require a rather long extrapolation in the light quark masses to their physical values.

The necessary guidance to perform the extrapolation is usually provided by chiral perturbation theory (\chpt) \cite{Gasser:1983yg,Gasser:1984gg}. This low-energy effective theory for QCD predicts the quark mass dependence of various physical quantities. A well-known example is the one-loop expression for the pion mass ($N_{f}=2$ and $m_{u}=m_{d}=m$), 
\bea\label{MassCHPT}
\frac{M_{\pi}^{2}}{2Bm} = 1 + \frac{2Bm}{32\pi^{2}f^{2}} \ln\frac{2Bm}{\Lambda^{2}} + \mbox{analytic}. 
\eea
In order to use this expression  one must make sure that one is in the chiral regime where \chpt\ holds. It is widely believed that a non-trivial check for this is provided by the logarithmic quark mass dependence in eq.\ \pref{MassCHPT}: Once the lattice data shows the characteristic curvature of the chiral logarithm one can apply eq.\ \pref{MassCHPT} with  confidence for the chiral extrapolation \cite{Bernard:2002pd}.

There is a potential problem with this argument. The derivation of \chpt\ is essentially based on symmetry properties of {\em continuum} QCD. Hence the continuum limit has to be taken first before \chpt\ can be employed to perform the chiral extrapolation.

There are various reasons why one may like to reverse this order. Obviously, as long as data for only one lattice spacing is available the continuum limit cannot be taken. Performing the chiral extrapolation first is also simpler in practice. 

Whatever the reasons might be, performing the chiral extrapolation before taking the continuum limit raises the question whether it is legitimate to use expressions derived in continuum \chpt. Optimistically one may hope to commit just a small error, assuming the lattice spacing is small. However, it is a priori not clear whether the functional form in eq.\ \pref{MassCHPT} is valid at all at non-zero lattice spacing.
This concern is even more justified taking into account that each of the traditional lattice fermions (Wilson and staggered) compromises chiral symmetry in some respects. In case eq.\ \pref{MassCHPT} is not appropriate at non-zero $a$, the question is which expression should be used instead.

\chpt\ can be formulated for lattice QCD at non-zero lattice spacing. The main idea goes back to two papers \cite{Sharpe:1998xm,Lee:1999zx} published about five years ago. Since then we have learned a lot about the chiral limit at non-zero lattice spacing. Moreover, formulae for masses, decay constants etc.\ were derived that include explicitly the contributions due to a non-vanishing lattice spacing. These formulae are the proper expressions one should use when the chiral extrapolation is performed before the continuum limit is taken. 

In this review I give an overview of \chpt\ at non-zero lattice spacing. I focus on the methodology, point out important differences compared to continuum \chpt\ and cover the main theoretical results. 

Some of these results entered already the analysis of numerical lattice data. I briefly comment on these analyses with the question in mind whether the simulations are carried out in the chiral regime so that \chpt\ can be applied. I do not discuss the physical results of these simulations for the hadron spectrum, quark masses and heavy quark physics. For these  I refer to the plenary talks given by K-I.\ Ishikawa, P.\ Rakow and M.\ Wingate at this conference \cite{IshikawaetalLat04}.

%
\section{\chpt\ for lattice theories}
%
The basic strategy for constructing \chpt\ for lattice theories at non-zero lattice spacing is a two-step matching to effective theories \cite{Sharpe:1998xm,Lee:1999zx}. We first write down Symanzik's effective theory \cite{Symanzik:1983dc,Symanzik:1983gh}, an effective continuum theory which describes the lattice theory close to the continuum limit. The cut-off effects  appear in terms of higher dimensional operators in the effective action and the effective operators, multiplied by powers of the lattice spacing $a$.
In the second step one derives the chiral Lagrangian for this effective theory using the standard arguments of \chpt. This results in a chiral expansion in which the dependence on the lattice spacing is made explicit.

The main r\^{o}le of Symanzik's effective theory in this two-step procedure is that it provides a systematic expansion of the lattice theory around continuum limit. It organizes the non-zero lattice spacing effects in powers of $a$ and therefore according to their importance when the continuum limit is approached. The structure of the higher dimensional operators in the Symanzik action determines if and how the cut-off effects break the symmetries of the corresponding continuum theory. In particular, the way chiral symmetry is broken by the lattice spacing effects is made transparent, which is crucial for constructing the chiral Lagrangian. Finally, Symanzik's effective theory is a continuum theory, and the well-established derivation of \chpt\ from continuum QCD can be readily extended to this effective theory with additional symmetry breaking parameters.   
%
\subsection{\chpt\ for Wilson fermions}
%
Consider lattice QCD with Wilson fermions.
Based on locality and the symmetries of the lattice theory, Symanzik's effective action is of the form \cite{Sheikholeslami:1985ij,Luscher:1996sc}
\bea
  S_{\rm Sym}  &=  &S_0  + aS_1  +  a^{2} S_2 + \ldots,\label{SymActionWF}\\ 
  S_k  &=& \sum\limits_i {c_i^{(k+4)} O_i^{(k+4)} },
\eea
where the $O_i^{(n)}$ are local operators of dimension $n$ constructed from the gauge and quark fields and their derivatives. The constants $c_i^{(n)}$ are unknown coefficients. The first term $S_{0}$ is the usual continuum QCD action. Note that the quark mass in the fermion part of $S_{0}$ includes the additive mass renormalization proportional to $1/a$, otherwise a term $S_{-1}$ would be present in eq.\ \pref{SymActionWF}. 

Using equations of motion there is essentially only the Pauli term in $S_{1}$,
\bea\label{PauliTerm}
S_{1} &=&  c_{1}  \int d^{4} x\,\overline \psi i\sigma _{\mu \nu } G_{\mu \nu } \psi.
\eea
This term breaks chiral symmetry, and its presence is a consequence of the explicit chiral symmetry breaking by the Wilson term in Wilson's fermion action. The complete list of dimension six operators in $S_{2}$ can be found in Ref.\ \cite{Sheikholeslami:1985ij}. Among the terms with fermions (bilinears and 4-quark-operators) are  operators which break chiral symmetry and ones which preserve it. It is also at this order in the Symanzik action that the lattice structure of the underlying theory shows up in form of quark bilinears that break O(4) rotation symmetry.

It should be mentioned that not all $a$ dependence is explicit in eq.\ \pref{SymActionWF}. The coefficients $c_i^{(n)}$ are functions of the gauge coupling $g^{2}$ and are therefore expected to show a presumably weak, logarithmic $a$ dependence.

Since the leading term in eq.\ \pref{SymActionWF} is the continuum QCD action we expect the lattice theory to exhibit the same spontaneous symmetry breaking pattern as in the continuum, provided both $m$ and $a$ are small. In that case the low-energy physics is dominated by pseudo Goldstone bosons, which acquire a non-zero mass due to the explicit chiral symmetry breaking by the quark mass and by the additional chiral symmetry breaking terms in $S_{1}$ and $S_{2}$. 

The low-energy chiral effective theory for these bosons, often called Wilson \chpt, is defined by a chiral effective Lagrangian. In order to construct this Lagrangian one writes down the most general Lagrangian that is invariant under the symmetries of the underlying Symanzik theory. Symmetry breaking terms are consistently included performing a spurion analysis. This procedure is analogous to the way the quark mass is included in continuum \chpt. Here, however, one has to perform a spurion analysis for each symmetry breaking term in eq.\ \pref{SymActionWF}, also those stemming from the discretization effects.

The Pauli term \pref{PauliTerm} is a particularly simple example for this procedure because it breaks chiral symmetry exactly like a mass term.
As usual, the chiral Lagrangian is parameterized in terms of $\Sigma = \exp(2i\Pi/f)$ with $\Pi$ being the matrix of Goldstone boson fields, which transforms under chiral transformations as $\Sigma\rightarrow L\Sigma R^{\dagger}$.  The ${\cal L}_{2}$-Lagrangian, containing the terms of ${\rm O}(p^{2},m,a)$, is found to be given by \cite{Rupak:2002sm}
\bea\label{WChPTL2}
\mathcal{L}_2 & =& 
\frac{{f^2 }}{4}\left\langle \partial_{\mu} \Sigma\partial_{\mu} \Sigma^\dag
\right\rangle 
- \frac{{f^2B }}{2} \left\langle m \Sigma^\dag + \Sigma
  m^{\dagger}\right\rangle \nn\\
  & & \hspace{1.8cm} -\frac{{f^2\tilde{W}_{0}}}{2}c_{1}a \left\langle\Sigma^\dag+\Sigma\right\rangle.   
\eea
The angled brackets denote traces over the flavor indices. The first line  contains the familiar terms from continuum \chpt, the kinetic and the mass term (here $m$ stands for the quark mass {\em  matrix}), multiplied by unknown low-energy constants $f$ and $B$. These two terms stem from the leading part $S_{0}$ in eq.\ \pref{SymActionWF}. The third term has its origin in the Pauli term, and it has, as expected, the structure of a degenerate mass term (degenerate because the Pauli term is diagonal in flavor space).  The coefficient $\tilde{W}_{0}$ is another low-energy constant not determined by symmetries. In contrast to $B$ the mass dimension of $\tilde{W}_{0}$ is three instead of one.

Since both  $\tilde{W}_{0}$ and $c_{1}$ are unknown parameters it is customary to combine them in form of one coefficient $W_{0}$. In this parameterization, however, the coefficient $W_{0}$ inherits the weak $a$ dependence of $c_{1}$ and is no longer a true constant. Furthermore, $W_{0}$ vanishes if the underlying lattice theory is non-perturbatively {\rm O}($a$) improved, because $c_{1}$ is zero in this case.

The $\mathcal{L}_4$-Lagrangian comprises all terms of ${\rm O}(p^{4},p^{2}m,m^{2}, p^{2}a,ma,a^{2})$ and it is of the form
\be\label{L4Lagrangian}
{\cal L}_{4}\,=\,{\cal L}^{{\rm GL}}_{4}(p^{4},p^{2}m,m^{2}) +  {\cal L}^{a}_{4}(p^{2}a,ma, a^{2}).
\ee
The first term on the r.h.s.\ is the well-known Gasser-Leutwyler Lagrangian
 \cite{Gasser:1983yg,Gasser:1984gg} 
stemming from the continuum part  in Symanzik's effective action. The second term parameterizes the additional chiral symmetry breaking effects coming from $S_{1}$ and $S_{2}$  \cite{Rupak:2002sm,Bar:2003mh}. It turns out that the operators in ${\cal L}^{a}_{4}$ are easily obtained from the Gasser-Leutwyler Lagrangian: Take any operator containing the mass matrix $m$ and replace it by $a$, this gives all terms in  ${\cal L}^{a}_{4}$. This simple final result is not obvious. Some 4-quark operators in $S_{2}$ break chiral symmetry in a different way than a mass term and rotational O(4) symmetry is broken at ${\rm O}(a^{2})$. However, the spurion analysis shows that all these effects do not enter the $\mathcal{L}_4$-Lagrangian, but only appear at higher orders in the chiral expansion \cite{Bar:2003mh}.

The total number of unknown low-energy constants in ${\cal L}_{4}$ is eighteen. Ten of those are  Gasser-Leutwyler coefficients in the Gasser-Leutwyler Lagrangian, while the lattice spacing effects contribute eight  additional unknown coefficients. This number is reduced to three for fully {\rm O}($a$) improved Wilson fermions, since all $am$ terms in ${\cal L}^{a}_{4}$ vanish in this case.

The main motivation for constructing a chiral effective Lagrangian for lattice   QCD is to compute the explicit $a$ dependence of observables and to guide the chiral extrapolation of numerical lattice data. Obviously, too many unknown low-energy constants limit the predictability of the chiral extrapolation. However, the situation is not as bad as the number eighteen may suggest. The number of free parameters in the chiral expressions for $m_{\pi}$ and $f_{\pi}$ is much smaller because many low-energy constants appear in particular linear combinations and can therefore be combined in form of a few unknown parameters. Still, an increased number of free parameters is the price one has to pay when one wants to perform the chiral extrapolation before taking the continuum limit.
  
Having constructed the chiral effective Lagrangian we can compute expressions for the pseudo scalar masses, decay constants, scattering lengths etc. However, in order to correctly describe the underlying lattice theory we need to properly match the parameters in both theories, which is not entirely straightforward. 

Starting from eq.\ \pref{WChPTL2} one  easily derives the tree-level expression ($m_{u}=m_{d}=m$ for simplicity) 
\bea\label{MpiTL}
M^{2}_{\pi} & =& 2 B m + 2 W_{0} a 
\eea
for the pion mass. Hence, the leading {\rm O}($a$) effect is a shift in the pion mass. Consequently, the pion mass does not vanish for $m = 0$. 

The mass $m$, however, is not the one that is usually used in the lattice theory \cite{Sharpe:1998xm,Aoki:2003yv}. 
Due to the explicit chiral symmetry breaking, the quark mass receives an additive mass renormalization proportional to $1/a$. A common definition for the renormalized quark mass is in terms of a vanishing pion mass. By definition, $M^{2}_{\pi} =0$ for $m^{\prime} = Z_{m}(m_{0} - m_{{\rm cr}})/a\,=\,0,$
where $m_{0}$ is the bare lattice mass. So defined, the critical quark mass $m_{\rm cr}$ accounts not only for the divergent additive mass shift, but also for finite shifts proportional to powers of $a$. Therefore, at leading order in the effective theory, the appropriate mass parameter is given by 
\bea\label{DefShiftedMass}
m'&=& m + a W_{0}/B.
\eea
Eq.\ \pref{MpiTL} now reads $M^{2}_{\pi}  = 2 B m'$ and the pion mass vanishes for $m'=0$, as required. Note that the proper parameter matching needs to be adjusted when we work beyond LO: The terms of ${\rm O}(a^{2})$ in ${\cal L}^{a}_{4}$ cause an additional shift in the critical mass and the r.h.s.\ of eq.\ \pref{DefShiftedMass} receives an additional contribution of ${\rm O}(a^{2})$.

Having found the proper parameter matching \pref{DefShiftedMass}, one can replace $m$ by $m'$ in the effective Lagrangian. After the replacement the {\rm O}($a$) term in ${\cal L}_{2}$ has disappeared, but the terms linear in $a$ in  ${\cal L}_{4}^{a}$ are still present. 

There are other definitions for the renormalized quark mass on the lattice. For example, one can define it in terms of the quark mass that enters the PCAC relation. All these definitions differ by ${\rm O}(a^{n})$ terms in the critical quark mass. Depending on the definition in the lattice theory the parameter matching might be different from eq. \pref{DefShiftedMass} and needs to  be done accordingly.

Another subtlety in Wilson \chpt\ has its origin in the presence of two expansion parameters, $m$ and $a$, or, to be more precise
\bea
\frac{2Bm'}{(4\pi f)^2}, &\quad & \frac{2W_{0}a}{(4\pi f)^2}.
\eea
Both parameters must be smaller than one for the chiral  expansion to make sense. But even if this requirement is satisfied, the {\em relative} size of these parameters is crucial for the proper power counting. In order to discuss this let us consider the following two terms which appear in the chiral Lagrangian:  
\bea
O_{1} & = & c_{1}m' \,\langle \Sigma + \Sigma^{\dagger}\rangle,\label{O1}\\
O_{2} & = & c_{2}a^2\langle \Sigma + \Sigma^{\dagger}\rangle^2.\label{O2}
\eea
$O_{1}$ is just the mass term in ${\cal L}_{2}$, parameterized in terms of $m'$ and using the short-hand notation $c_{1} = f^{2}B/4$. $O_{2}$ appears in ${\cal L}_{4}^{a}$ and the coefficient $c_{2}$ denotes a particular combination of low-energy constants in ${\cal L}_{4}^{a}$. 

As long as $c_{1}m' \gg c_{2} a^2$ the term $O_{2}$ is much smaller than $O_{1}$ and can safely be considered a next-to-leading order (NLO) contribution. However, decreasing the quark mass at fixed lattice spacing (this is approximately done in numerical lattice simulations at fixed $\beta$) one will eventually enter a regime where both terms are of comparable size. In this regime both contributions should be taken to be of leading order (LO). 

The regime $c_{1}m' \gg c_{2} a^2$ is considered in Ref.\ \cite{Bar:2003mh}, and  the pseudo-scalar mass was calculated to one loop as an example. The resulting expression is essentially the one-loop continuum expression, containing the non-analytic continuum chiral logarithms, plus additional analytic terms proportional to $am'$ and $a^{2}$. 

Qualitative changes start to occur in the regime $c_{1}m' \approx c_{2} a^2$. To be more concrete let us consider the leading terms in the potential energy for two degenerate flavors \cite{Sharpe:1998xm},
\be\label{potential}
V = - \frac{c_{1}}{4}m' \,\langle \Sigma + \Sigma^{\dagger}\rangle + \frac{c_{2}}{16}a^2\langle \Sigma + \Sigma^{\dagger}\rangle^2,
\ee
which is essentially the sum of the terms in eqs.\ \pref{O1} and \pref{O2} (the relative sign is convention).\footnote{Note that the definitions for $c_{1}$ and $c_{2}$ differ from Ref.\ \cite{Sharpe:1998xm}. I have pulled out the factors $m'$ and $a^{2}$.} For $c_{1}m' \approx c_{2} a^2$ the two terms in the potential are of comparable size, and the competition between them can result in a non-trivial ground state. 

It turns out that there are only two different scenarios possible, and the sign of $c_{2}$ determines which of those is realized \cite{Sharpe:1998xm}. If $c_{2}$  is positive, the ground state configuration $\Sigma_{0}$ is no longer proportional to the identity for $m'< 2c_{2}a^{2}/c_{1}$. Parity and flavor are spontaneously broken and massless pions exist even at non-zero lattice spacing. In other words, the effective theory predicts the properties of the Aoki phase, which was proposed a long time ago \cite{Aoki:1983qi}. The alternative scenario with negative $c_{2}$ exhibits a first order phase transition where $\Sigma_{0} = 1$ changes sign. Parity and flavor are unbroken irrespective of the size of $m'$ and no massless pions exist at non-zero $a$.

The same analysis for quenched Lattice QCD is more subtle due to the ghost fields and the graded symmetry group. Nevertheless, the conclusion is that the phase structure is the same as in the unquenched theory \cite{SharpeLat2004}.  

The chiral effective theory cannot predict whether $c_{2}$ is positive or negative. After all, $c_{2}$ is a combination of unknown low-energy constants whose values are essentially determined by the action of the underlying lattice theory. In particular, magnitude and sign of $c_{2}$ can be different for improved Wilson fermions. 

Numerical data to date support the existence of an Aoki phase for quenched simulations \cite{Aoki:1995ft,Aoki:1997fm}. Recent  unquenched  2-flavor simulations using the plaquette gauge action and unimproved Wilson fermions at $\beta = 5.2$ show evidence for a first-order phase transition \cite{FarchioniLat2004,Farchioni:2004us} (see also Ref.\ \cite{Ilgenfritz:2003gw}). The results suggest that  the scenario with negative $c_{2}$ is realized for this particular lattice action, but more data is needed to draw a definite conclusion. 
In the scenario with negative $c_{2}$ the minimal pion mass is determined by $|c_{2}|$. 
Hence, from the point of view of numerical  simulations, $|c_{2}|$ should be as small as possible, and it is an open question which lattice action is most suitable in this respect. 

A second feature of the regime $c_{1}m' \approx c_{2} a^2$ is that additional chiral log contributions appear in the one-loop expressions for observables. This has been shown for the two flavor case in Ref.  \cite{Aoki:2003yv}. The ${\rm O}(a^{2})$ term in \pref{O2} and also the ${\rm O}(a m')$ contributions are kept at LO in the chiral Lagrangian. These terms give rise to additional vertices proportional to $a^{2}$ and $am'$ and therefore to additional loop diagrams. Explicitly, the one-loop expression for the pion mass is given by
\bea\label{1loopAoki}
 \frac{M_{\pi}^2}{2Bm'}&
        =& 1 + \frac{m'(2B +w_{1} a)}{32\pi^2 f^2} \ln\frac{2Bm'}{\Lambda^{2}} \nn\\ 
        & & \!+ \frac{w_{0} a^{2}}{32\pi^2 f^2}\ln\frac{2Bm'}{\Lambda^{2}} + \mbox{analytic}. 
\eea
Here $w_{0}$ and $w_{1}$ denote some combinations of unknown low-energy constants and $m'$ includes the ${\rm O}(a^{2})$ shift coming from the $c_{2}a^{2}$ term in the potential \pref{potential}.

Eq.\ \pref{1loopAoki} coincides with the continuum one-loop expression in eq.\ \pref{MassCHPT} in the limit $a\rightarrow 0$. However, the coefficient of the $m'\ln m'$ term receives a correction of {\rm O}($a$). Furthermore, the lattice spacing effects generate an additional $a^{2}\ln m'$ contribution. 
  
This $a^{2}\ln m'$ contribution will eventually become dominant when we decrease $m'$ further. In fact, the r.h.s.\ of eq.\ \pref{1loopAoki} diverges in the $m'\rightarrow 0$ limit ($M_{\pi}^2$ itself, however, remains finite). 
Toward the chiral limit terms proportional to $a^{2}(\ln 2Bm')^{n},\, n=2,3,\ldots$ become more and more important. Aoki performed a resummation of these terms and derived a resummed one-loop formulae for the pion mass:
\bea\label{1loopAokiResummed}
 \frac{M_{\pi}^2}{2\tilde{B}m'}&
        =& \left[1 + \frac{(2B +w_{1} a)m'}{32\pi^2 f^2} \ln\frac{2Bm'}{\Lambda^{2}}\right]  \nn\\ 
        & & \times \left\{\ln \frac{2Bm'}{\tilde{\Lambda}^{2}}\right\}^{\tilde{w}_{0}a^{2}/32\pi^{2}f^{2}}\hspace{-1.3cm} + \mbox{ analytic}. 
\eea
Expanding $\{\ldots\}^{a^{2}\ldots}$ and dropping higher powers of  $a^{2}(\ln 2Bm')^{n}$ one recovers eq.\ \pref{1loopAoki}. 

The derivation of eq.\ \pref{1loopAokiResummed} assumes the Aoki scenario for the phase diagram (positive $c_{2}$) where the pion becomes massless at a second order phase transition point. Approaching this point the correlation length (or the inverse pion mass) diverges with the critical exponent of a four dimensional scalar theory. Comparing the general form for the diverging correlation length with eq.\ \pref{1loopAoki} one can match the parameters and obtains formula \pref{1loopAokiResummed}. Note that the parameters with a tilde in eq.\ \pref{1loopAokiResummed}
may be different from those in eq.\ \pref{1loopAoki}. The matching at this order does not determine the parameters unambiguously \cite{Aoki:2003yv}.

Analogous one-loop calculations for the pseudo-scalar decay constant and the PCAC quark mass $m_{{\rm AWI}}$ were also carried out \cite{Aoki:2003yv}. The results exhibit the same qualitative features as in  eqs.\ (\ref{1loopAoki}) and (\ref{1loopAokiResummed}). Additional logarithmic contributions proportional to $am'\ln m'$ and $a^{2}\ln m'$ are present and modify the familiar results obtained in continuum \chpt.

These results show that the chiral limit at non-zero lattice spacing is quite different from the one 
in continuum \chpt. The differences become more pronounced the smaller the mass $m'$ is.
This is a warning that expectations concerning the quark mass dependence of $M_{\pi}^{2}, f_{\pi}$ and other quantities based on continuum \chpt\ might be misleading when naively applied to lattice QCD. Performing the chiral extrapolation of lattice data using the chiral forms of continuum \chpt, as is often done, might not be justified. 

From a practical point of view the crucial question is what $c_{1}m' \approx c_{2} a^2$ precisely means.
This question is not easily answered, since nothing is known about the size of the low-energy constants  that go into $c_{2}$. A simple dimensional analysis tells us that $c_{1}$ and $c_{2}$ are of mass dimension three and six, respectively. Hence  $c_{1}m' \approx c_{2} a^2$ can be translated into $m' \approx a^2\Lambda_{\rm QCD}^{3}$, assuming that the size of any dimensionful quantity is determined by the typical QCD scale. 
This argument should be taken with care, since factors of 2 or 3 are easily amplified by taking powers.

Nevertheless, assuming $\Lambda_{\rm QCD}\approx 300$MeV and a lattice spacing $a \approx 0.15$fm we find $a^2\Lambda_{\rm QCD}^{3} \approx 15$MeV. Even though the physical quark masses simulated in present day numerical simulations are larger than this value, they are probably not large enough to conclude $m' \gg a^2\Lambda_{\rm QCD}^{3}$ and to neglect the effects due to a non-zero $a$. Ultimately the fits of the chiral forms to the numerical lattice data have to decide which power counting is more appropriate in explaining the data at hand.
%
\subsection{Partially quenched and mixed fermion theories}
%
The construction of a chiral effective Lagrangian is readily extended to partially quenched lattice QCD with different masses for the sea and valence Wilson fermions. The partially quenched lattice theory is described by a lattice action with sea, valence and ghost quarks. The Symanzik action through ${\rm O}(a^{2})$ is obtained as in the unquenched case, based on locality and the symmetries of the lattice theory \cite{Bar:2003mh}. The chiral effective Lagrangian through ${\rm O}(a^{2})$ has the same form as in the unquenched case, with the angled brackets (cf.\ eq.\ \pref{MpiTL}) now representing super-traces and the field $\Sigma$ reflecting the larger flavor content of partially quenched \chpt. 

Mixed fermion (or hybrid) theories are a generalization of partially quenched lattice theories. In addition to choosing different quark masses, the lattice Dirac operator is different in the sea and valence sector. 
Particularly interesting combinations contain a Dirac operator for the sea quarks that is fast to simulate, i.e.\ staggered or (twisted mass) Wilson fermions, and Ginsparg-Wilson fermions for the valence quarks, realized by domain-wall \cite{Kaplan:1992bt} or overlap fermions \cite{Narayanan:1992wx,Narayanan:1993sk}. This type of mixed fermion simulations offers an efficient compromise towards full unquenched simulations with Ginsparg-Wilson fermions. Some new results using configurations generated  with staggered sea quarks and domain wall or overlap valence quarks have been reported at this conference \cite{FlemingLat2004}.

The naive argument why mixed fermion theories are expected to give meaningful physical results is that 
the two Dirac operators differ by terms of {\rm O}($a$) and these should vanish in the continuum limit. However, there are potential dangers. Unitarity is lost and it is restored in the continuum limit only. This is in contrast to partially quenched theories with the same Dirac operator, which become unitary when the valence and sea quarks are chosen equal. Moreover, it is not at all obvious that a ``better'' Dirac operator for the valence quarks (one that has exact chiral symmetry at non-zero $a$) automatically implies better results for physical quantities. Analytic control of the expected 
 {\rm O}($a$) difference is clearly desirable. 
 
The chiral effective Lagrangian for lattice QCD with Wilson sea quarks and Ginsparg-Wilson valence quarks was constructed in \cite{Bar:2002nr,Bar:2003mh}. It turns out that the effective Lagrangian contains one more operator at ${\rm O}(a^{2})$ and consequently one more unknown low-energy constant than the Lagrangian for Wilson sea and Wilson valence quarks. Nevertheless, an explicit calculation shows that this additional unknown constant does not enter the one-loop result for the pion mass of a pion made of two valence quarks. In fact, compared to the case with Wilson sea and Wilson valence quarks one finds a reduced $a$ dependence.

This example demonstrates that the cut-off dependence of mixed fermion theories can be studied analytically using the chiral effective field theory. Work on the mixed theory with staggered sea and Ginsparg-Wilson valence quarks is in progress \cite{BarLat20042}.

\subsection{Twisted mass Lattice QCD}

The advantages of the twisted mass formulation of lattice QCD [tmLQCD] \cite{Frezzotti:2001ea,Frezzotti:2000nk,Frezzotti:2002iv} with Wilson fermions have been reviewed by R.\ Frezzotti at this conference \cite{FrezzottiLat2004}. A twisted mass term
\bea\label{twistedmass}
m_{\rm tm}& = & m + i \mu \gamma_{5}\sigma_{3}
\eea
 protects the Wilson-Dirac operator against very small eigenvalues and solves the problem of exceptional configurations. Recent results indicate that unquenched simulations also benefit from a twisted mass, and their ``numerical cost'' is comparable with staggered fermions \cite{KennedyLat2004}.  Hence, simulations with smaller physical quark masses seem possible in tmLQCD and the chiral regime is probably easier reached than with an untwisted mass term.
 In addition, certain physical quantities like hadron masses are automatically {\rm O}($a$) improved \cite{Frezzotti:2003ni,Frezzotti:2004wz}.

The construction of the chiral effective theory for tmLQCD follows the same two-step procedure that was described before \cite{WuLat04,Sharpe:2004ps}. The form of Symanzik's effective action is as in eq.\ \pref{SymActionWF}, with the leading term $S_{0}$ now being the continuum twisted mass QCD action. Since a twisted mass term breaks parity and flavor, there are more terms present in $S_{1}$ and $S_{2}$ compared to the untwisted case. Nevertheless, after performing the spurion analysis one finds the same ${\cal L}_{2}$- and ${\cal L}_{4}$-Lagrangian as in eq.\ \pref{WChPTL2} and \pref{L4Lagrangian}, with $m$ replaced by the twisted mass $m_{\rm tm}$.
 
The effective Lagrangian was used to analyze the phase diagram of tmLQCD, generalizing the analysis for the untwisted theory \cite{Munster:2004am,WuLat04,MunsterLat2004,Sharpe:2004ps,Scorzato:2004da}. As before, there exist two scenarios, depending on the sign of (the same) coefficient $c_{2}$, and for a vanishing mass $\mu$ one recovers the results in the untwisted case.

One-loop calculations for $M_{\pi}^{2}$ and $f_{\pi}$ have been performed  for the regime $\mu \gg a^{2} \Lambda_{{\rm QCD}}^{3}$, but only the terms linear in $a$ were kept \cite{Munster:2003ba,Munster:2004dj}. It is desirable to repeat these calculations including the ${\rm O}(a^{2})$ terms and with a power counting appropriate for $\mu \approx a^{2} \Lambda_{{\rm QCD}}^{3}$, since numerical simulations with small twisted mass may well be in this regime.

The masses of the neutral and charged pions differ due to explicit flavor breaking by a twisted mass term. The mass splitting is found to be of ${\rm O}(a^{2})$ and proportional to $c_{2}$ \cite{WuLat04,Sharpe:2004ps,Scorzato:2004da},
\bea
M_{\pi_{3}}^{2} - M_{\pi_{\pm}}^{2}  &=& \frac{2c_{2}}{f^{2}} a^{2} (1 - \cos^{2} \phi),
\eea
where the angle $\phi$ parameterizes the vacuum state $\Sigma_{0} = \exp i\phi\tau_{3}$ of the effective theory. Hence, as pointed out in \cite{Scorzato:2004da}, the sign of $c_{2}$ can be determined, at least in principle, by measuring the mass difference of the charged and neutral pions.

A proof for automatic {\rm O}($a$) improvement at maximal twist was presented in Ref.\ \cite{Frezzotti:2003ni}.
It was subsequently shown \cite{Aoki:2004ta} that a crucial assumption about the critical quark mass does not hold if $c_{2}$ is positive. Consequently, {\rm O}($a$) improvement is lost unless $m\gg a^{2}\Lambda_{\rm QCD}^{3}$.  This restriction on the quark mass, however, can be avoided with a different definition for maximal twist, and {\rm O}($a$) improvement can be guaranteed irrespective of the size of the quark mass. 

The differences between different definitions for maximal twist was illustrated using the framework of the chiral effective theory. 
The absence or presence of the leading O$(a)$ effect in the pion mass, depending on the definition for maximal twist and the size of the quark mass, is explicitly shown in Ref.\ \cite{Aoki:2004ta}. 
%
\subsection{Nucleon properties}
%
Starting from the Symanzik action in eq.\ \pref{SymActionWF}, continuum Baryon \chpt\ has been extended to accommodate the leading linear $a$ dependence due to the Pauli term $S_{1}$ \cite{Beane:2003xv}. This extension is rather straightforward. Since the Pauli term breaks chiral symmetry like a mass term,  the construction of the chiral effective Lagrangian involves one additional spurion field proportional to $a$, but is otherwise analogous to the construction based on continuum QCD \cite{Jenkins:1990jv}. 

Assuming a power counting with $m \approx a\Lambda_{\rm QCD}^{2}$, a variety of nucleon properties (masses, magnetic moments, matrix elements of the axial vector current etc.) have been computed in the one-loop approximation. At this order the main effect of the non-zero lattice spacing is the shift of the pseudo scalar masses in eq.\ \pref{MpiTL}. As discussed before, this shift might already be absorbed in the definition of the critical mass. Non-trivial effects, however, can be expected at ${\rm O}(a^{2})$.  

Electromagnetic properties of baryons and mesons (charge radii, magnetic moments etc.) including the linear lattice spacing contribution have also been discussed in Ref.\ \cite{Arndt:2004we}. Again, the main $a$ effect is implicit in the pseudo-scalar masses, for example for the charge radius of the $\phi$ meson. It would be interesting to extend these results by including the ${\rm O}(a^{2})$ corrections.
%
\subsection{\chpt\ for staggered fermions }
%
Staggered fermions are numerically very fast to simulate compared with other lattice fermions. They possess an exact axial U(1) symmetry at non-zero lattice spacing, which protects the quark mass from an additive renormalization. As a result, lattice QCD simulations with staggered fermions reach significantly smaller values for $M_{\pi}/M_{\rho}$ than those using Wilson fermions. The numerical performance of the known lattice fermions in unquenched simulations was reviewed by A.\ Kennedy at this conference \cite{KennedyLat2004}.
The major disadvantage of staggered fermions is  that they do not solve the fermion doubling problem completely: Each flavor comes in four different tastes.

In order to reduce the number of tastes one usually employs the so-called ``fourth root trick'': The fermion determinant of the staggered Dirac operator is replaced by $\sqrt[4]{\det D}$ in numerical lattice simulations. This trick legitimately raises the question whether the fourth root theory correctly describes QCD in the continuum limit. By taking the fourth root one sacrifices the locality of the theory and all known universality arguments no longer hold. This and additional problems are reviewed in Ref.\ \cite{Jansen:2003nt}.

Using the fourth root trick also poses  a problem for constructing a chiral effective theory. Since the lattice theory is no longer local, it is not described by a local Symanzik theory close to the continuum limit. Hence the previously described two-step procedure cannot be applied directly.

To circumvent this problem the following strategy has been proposed \cite{Bernard:2001yj}. One first considers lattice QCD with $N_{f}$ staggered flavors without using the fourth root trick. This theory is local and one can indeed construct Symanzik's effective theory. The leading term $S_{0}$ is the continuum QCD action with $4N_{f}$ fermions (4 tastes for each flavor). Assuming spontaneous chiral symmetry breaking one constructs the chiral Lagrangian for this lattice theory with $(4N_{f} )^{2} - 1$ pseudo Goldstone bosons. Starting from this Lagrangian one calculates pseudo scalar masses, decay constants etc.\ to the desired order (one loop in practice). These results are finally corrected for taking the fourth root of the determinant. This adjustment amounts to properly placing factors of 1/4 for each sea quark loop contribution. This step requires, besides performing a partially quenched calculation in order to distinguish between sea and valence quarks, that the meson diagrams in the effective theory are correctly interpreted in terms of the underlying quark diagrams \cite{Sharpe:1992ft}.

This procedure, staggered \chpt\ for short, is field theoretically not absolutely rigorous. Potential non-local contributions due to taking the fourth root of the fermion determinant would not be captured by it. 
Consequently, the validity of the fourth root trick would be seriously questioned if staggered lattice data cannot be described by staggered \chpt. Turning this numerical argument around is not so simple. Even if no problems are found numerically, some doubts may still remain. More analytic studies are certainly desired (see Ref.\ \cite{AdamsLat2004} for a recent example).

Putting these issues aside, the construction of staggered \chpt\ follows the two-step procedure outlined before. In one respect staggered \chpt\ is  simpler than Wilson \chpt\  because the quark mass is not additively renormalized.

The form of Symanzik's effective action is as in eq.\ \pref{SymActionWF}, but the symmetries of the staggered lattice action exclude any terms of dimension 3 and 5, so $S_{1} $ vanishes \cite{Sharpe:1993ng,Luo:1997tt,Lee:1999zx}. The leading term $S_{0}$ is the continuum QCD action with $N_{f}$ flavors, each coming in four different tastes. $S_{0}$ possesses an exact SU(4) taste symmetry for each flavor, but this symmetry is broken at ${\rm O}(a^{2})$ by dimension six operators in $S_{2}$. In addition, SO(4) rotation invariance is broken at this order.

The chiral Lagrangian is constructed in the same way as described for Wilson fermions. The symmetry breaking terms are consistently included by performing  a spurion analysis. The generic form of the ${\cal L}_{2}$-Lagrangian is given by \cite{Lee:1999zx}
\bea\label{LOmassStaggered}
{\cal L}_{2} & =& {\cal L}_{\rm kin}+{\cal L}_{\rm mass}+ a^{2}{\cal V}.
\eea
The kinetic and the mass term are as in eq.\ \pref{WChPTL2}, however, the field $\Sigma$ and the mass matrix are $4N_{f}\times4N_{f}$ matrices, reflecting the larger particle content due to the taste degree of freedom.

The potential ${\cal V}= \sum_{i}c_{i}{\cal O}_{i}$ comprises eight taste symmetry breaking operators ${\cal O}_{i}$, each of which is multiplied by an unknown low-energy constant $c_{i}$ \cite{Lee:1999zx,Bernard:2001yj,Aubin:2003mg} (two of the operators are redundant in the one flavor case). However, the taste symmetry is not completely broken and ${\cal V}$ retains an accidental SO(4) taste symmetry. Moreover, ${\cal V}$ is SO(4) rotationally invariant, even though rotation invariance is broken in the Symanzik action at ${\rm O}(a^{2})$. 

Expanding eq.\  \pref{LOmassStaggered} to quadratic order in the pion fields one obtains ($m=m_{u}=m_{d}$)
\bea\label{LOMassesSChPT}
M_{\pi_{{i}}}^{2} & =& 2Bm + a^{2} \Delta(\xi_{i})
\eea
for the leading order pion mass where the index $i=5,\mu 5,\mu\nu,\mu,I$ labels the different tastes and the $\xi_{i}$ denote the SU(4) taste generators \cite{Lee:1999zx,Aubin:2003mg}. The accidental SO(4) taste symmetry  of ${\cal L}_{2}$ implies that the mass shift $\Delta(\xi_{i})$ is the same for all $\xi_{\mu}$, all $\xi_{5\mu}$, and all $\xi_{\mu\nu}$. The shift $ \Delta(\xi_{5})$ for the Goldstone pion $\pi_{5}$ is of course zero because of the exact axial U(1) symmetry.

The mass degeneracy is not exact since it is a consequence of the SO(4) taste symmetry of ${\cal L}_{2}$,  and this symmetry is broken by higher order terms  in the effective Lagrangian. Nevertheless, the approximate degeneracy is clearly observed in numerical lattice data, both in quenched \cite{Ishizuka:1993mt,Aoki:1999av} and unquenched simulations \cite{Orginos:1998ue,Aubin:2004fs}. In fact, it was observed long before the analysis in the chiral effective theory offered a theoretical understanding for it. 

The chiral effective theory does not say anything about the mass shifts $\Delta(\xi_{i})$, neither the sign nor the size. A negative shift would imply a vanishing meson mass before the chiral limit and the existence of non-trivial phases, similar to the Aoki phase for Wilson fermions. However, the mass shifts observed in numerical simulations are all positive. Moreover, the shifts are fairly large. The most recent simulations by the MILC collaboration \cite{Aubin:2004fs} still show significant mass shifts, even though the lattice spacing is fairly small ($a_{\rm min} \approx 0.09$fm) and the highly improved Asqtad quark action is used. In fact, the lattice spacing contribution
$a^{2}\Delta(\xi_{i})$ to the pion masses is of the same order as the quark mass contribution $2Bm$. This  justifies, even requires, to consider the contribution $a^{2}{\cal V}$ in the leading order Lagrangian, while all the terms of ${\rm O}(p^{2}a^{2}, ma^{2}, a^{4}) $ enter the next-to-leading order Lagrangian ${\cal L}_{4}$.

The full next-to-leading order Lagrangian has been constructed just recently \cite{vandewaterLat2004,Sharpe:2004is}. The number of terms is fairly large, more than 200 operators enter ${\cal L}_{4}$. At this order in the chiral expansion the accidental SO(4) taste symmetry of ${\cal L}_{2}$ is broken by terms of ${\rm O}(p^{2}a^{2})$, and the symmetry group of the effective theory coincides with the one of the underlying lattice theory.

Despite the fact that there are so many operators at NLO, some non-trivial predictions have been found \cite{vandewaterLat2004,Sharpe:2004is}. At NLO the SO(4) taste symmetry is essentially only broken by the ${\rm O}(p^{2}a^{2})$ terms, and these terms contribute to both the pseudo-scalar masses and the matrix elements of the pseudo scalar density, $\langle0|P^{a}|\pi_{b}\rangle = \delta^{ab} f_{\pi_{a}}^{P}$. 
It turns out that there are sufficiently many independent relations in order to predict relationships that do not contain any unknown low-energy constants. For example, one finds ($k=1,2,3$)
\bea\label{NTRelationfPMass}
\frac{f_{\pi_k}^{P} - f_{\pi_4}^{P}}{f_{\pi_{k}}^{P} + f_{\pi_{4}}^{P}} & =& \frac{1}{2} \frac{M^{2}_{\pi_k}-M^{2}_{\pi_4}}{M^{2}_{\pi_k} + M^{2}_{\pi_4}}, 
\eea
and the same relations involving the taste pairs $\pi_{k5}, \pi_{45}$ and $\pi_{lm}, \pi_{k5}, (k,l,m=1,2,3)$. These expressions show that the degeneracies among the tastes with $\xi_{\mu}, \xi_{\mu5}$ and $\xi_{\mu\nu}$ are removed.

Non-trivial predictions for the pion dispersion relations have also been established \cite{vandewaterLat2004,Sharpe:2004is}, but eq.\ \pref{NTRelationfPMass} seems to be the simplest prediction to test in numerical simulations. To this end it is extremely beneficial  that the renormalization factors entering eq.\ \pref{NTRelationfPMass} are SO(4) taste invariant. They are  therefore identical for both tastes and cancel in the ratio, hence permitting  eq.\ \pref{NTRelationfPMass} to be tested using bare lattice operators. This is not true in general. In particular it would not hold if the l.h.s.\ in eq.\ \pref{NTRelationfPMass} involved the matrix element  $\langle0|A_{\mu}^{a}|\pi_{a}\rangle$ of the axial vector current $A_{\mu}^{a}$. 

Checking these relationships in numerical simulations may serve as an additional test for the fourth root trick. Most of these relations are unaffected by the necessary modifications  due to taking the fourth root of the determinant. The trick is certainly not justified if the simulations cannot reproduce these predictions.

Taking the $a^{2}{\cal V}$ term in ${\cal L}_{2}$ to be of leading order  gives rise to additional interaction vertices and consequently to logarithmic contributions proportional to $a^{2}$ in loop calculations for physical quantities.
The masses and decay constants of the pseudo-Goldstone bosons have been computed to one loop in Refs.\ \cite{Aubin:2003mg,Aubin:2003uc}. For three degenerate flavors the mass of the charged  Goldstone pion reads \cite{Aubin:2003mg}
\bea\label{PSMassoneloopSChPT}
 \frac{M^2_{\pi^+_5}}{2Bm}
       &  = & 1 + \frac{1}{48\pi^2 f^2} M_{\pi_I}^2\ln\frac{M_{\pi_I}^2}{\Lambda^{2}} \nn\\
       & & \hspace{-1.3cm}+\, \frac{1}{12\pi^2f^2} \biggl[M^2_{\eta'_V}\ln\frac{M_{\eta'_{V}}^2}{\Lambda^{2}}- M^2_{\pi_V}\ln\frac{M_{\pi_{V}}^2}{\Lambda^{2}})\biggr]\nn\\
& &         \hspace{-1.3cm}+\, \frac{1}{12\pi^2f^2} \biggl[M^2_{\eta'_A}\ln\frac{M_{\eta'_{A}}^2}{\Lambda^{2}}- M^2_{\pi_A}\ln\frac{M_{\pi_{A}}^2}{\Lambda^{2}})\biggr]\nn\\[0.5ex]
        & &  \hspace{-1.3cm}+\mbox{ analytic terms.}
\eea
The dependence on the lattice spacing is implicit and enters in form of the leading order pseudo-scalar masses given in eq.\ \pref{LOMassesSChPT}. The first line reproduces the familiar continuum \chpt\ result for $a\rightarrow0$, while the second and third line vanish in this limit (note that all masses $M_{\pi_{{i}}}^{2} $ become degenerate in the continuum limit). At non-zero $a$, however, this result may differ significantly from the result in the continuum limit, depending on the size of the mass shifts for the various 
pseudo-scalar mesons. Note that the perhaps naively expected term proportional to $M_{\pi^+_5}^2\ln(M_{\pi^+_5}^2/\Lambda^{2})$ does not appear on the r.h.s.\ of eq.\ \pref{PSMassoneloopSChPT}.

The one-loop expressions for the decay constants $f_{\pi^{+}_{5}}$ and $f_{K^{+}_{5}}$ show similar qualitative modifications compared to the continuum \chpt\ results for these quantities \cite{Aubin:2003uc}. It should be mentioned that the expressions for 2+1 partially quenched flavors, which is the relevant case for the simulations carried out by the MILC collaboration, are much lengthier than the special example for 3 unquenched flavors in eq.\ \pref{PSMassoneloopSChPT}.

\subsection{Staggered \chpt\ for heavy-light mesons}

Staggered \chpt\ has been extended to describe heavy-light mesons \cite{AubinLat2004}. The starting point is Symanzik's effective action for staggered light quarks given in \pref{LOmassStaggered}. No Symanzik analysis was done for the heavy quark lattice action. Instead, the leading order HQET action is assumed to give a proper description of the heavy quark. This assumption neglects the discretization effects due to the heavy quark and is only justified for either a highly improved heavy quark lattice action or when lattice HQET \cite{Eichten:1989zv} is used. 

Starting from this effective continuum theory it is straightforward to generalize the arguments of continuum heavy-light \chpt\ \cite{Wise:1992hn} and construct the chiral Lagrangian. One difference is that we now have three expansion parameters: the quark mass $m$, the lattice spacing $a$ and the residual momentum $k$ of the heavy-light meson. 

Based on the chiral Lagrangian the decay constant $f_{B}$ was calculated to one loop. The additional vertices proportional to $a^{2}$ generate extra chiral logarithms. These terms can again significantly change the quark mass dependence that the continuum expressions predict. An illustrative plot showing this can be found in \cite{AubinLat2004}.

\section{Comparison with  numerical data}

The main motivation for constructing chiral effective Lagrangians for lattice theories is to capture the discretization effects analytically and to guide the chiral extrapolation of numerical lattice data without taking the continuum limit first. Of course, one has to be in the regime where the chiral effective theory provides a valid description of the lattice data. This can only be checked by trying to fit  the chiral fit forms to the lattice data. In case one obtains fits with a good $\chi^{2}$ and reasonable values for the unknown coefficients one gains confidence that the chiral effective theory describes the data. 

Both the CP-PACS collaboration \cite{Namekawa:2004bi,NamekawaLat2004} and the qq+q collaboration \cite{Farchioni:2003nf,Farchioni:2004tv,ScholzLat2004} used Wilson \chpt\  to analyze their unquenched lattice data which were obtained with Wilson fermions. 

\begin{figure}[t]
\vspace{-1mm}
\begin{center}
\scalebox{0.60}{\includegraphics{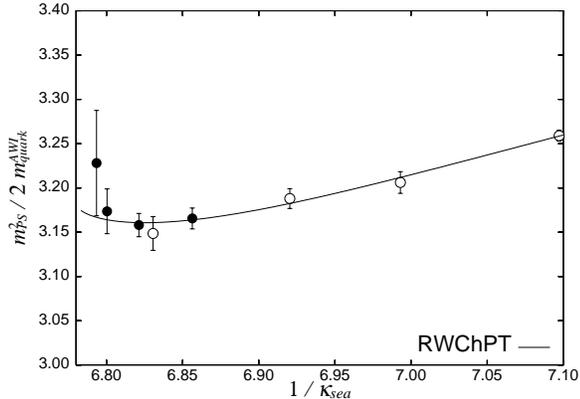}}
\end{center}
\vspace{-10mm}
\caption{Fit result for $M^{2}_{\pi}/2m_{\rm AWI}$ using resummed Wilson \chpt. From Ref.\ \cite{Namekawa:2004bi}.}
\label{figCPPACS1}
\vspace{-6mm}
\end{figure}

The CP-PACS collaboration generated their data with an RG-improved gauge action and a meanfield improved clover quark action with 2 flavors. The simulations were performed at one lattice spacing $a \approx 0.2$fm and for eight different quark masses corresponding to the range $ M_{\pi}/M_{\rho}=  0.35 -  0.8$. Data for four mass values were generated some time ago \cite{AliKhan:2001tx} and were combined with new data \cite{Namekawa:2004bi}.

Chiral fits for $M_{\pi}^{2}$ and $m_{\rm AWI}$ were performed using the results of continuum  \chpt\ as well as the Wilson \chpt\  expressions in eqn.\ \pref{1loopAoki} and \pref{1loopAokiResummed} for $M_{\pi}^{2}$ and the corresponding expressions for $m_{\rm AWI}$. The power counting underlying these formulae seems appropriate since the bare quark mass $am'$ is of ${\rm O}(a^{3}\Lambda_{\rm QCD}^{3})$ in the CP-PACS simulations.

Fig.\ \ref{figCPPACS1} shows the fit result for $M_{\pi}^{2}/2m_{\rm AWI}$ using the resummed formulae for both quantities. A good fit is obtained including all data points. 
Using the unresummed formulae (cf.\  eq.\ \pref{1loopAoki} for $M_{\pi}^{2}$) gives similar results.  
A reasonable fit is also possible with  the 1-loop continuum expression when the three heaviest data points are excluded from the fit. However, the lowest five data points can also be fitted by a straight line within errors. Hence, even though fits to continuum \chpt\ are possible there is  no clear evidence for the curvature due to the chiral logarithms. 

The good fit result over the whole range of quark masses is quite unexpected since the chiral expansion is not expected to work at such high values for $M_{\pi}/M_{\rho}$. 
The reason why the formulae of Wilson \chpt\  work so well in figure \ref{figCPPACS1} can be traced back to the coefficient of the $m'\ln m'$ term. The result for this coefficient is roughly 80$\%$ smaller than the value expected from continuum \chpt. Hence the curvature due to the chiral logarithm is highly suppressed and the fairly linear  data is fitted well. 

This strong suppression is slightly surprising. The coefficient of the $m'\ln m'$ term is proportional to $(2B - \tilde\omega_{1} a)$ where $\tilde\omega_{1}$ is the difference of the $\omega_{1}$ coefficients in the chiral expressions for $M_{\pi}^{2}$ and $m_{\rm AWI}$. Large linear lattice spacing effects with $\tilde\omega_{1} a = {\rm O}(2B)$ are required in order to achieve the 80$\%$ suppression. Since 
a meanfield improved quark action was used one would have expected smaller values for $\tilde\omega_{1}$.\footnote{$\tilde\omega_{1}$ would be zero for non-perturbatively {\rm O}($a$) improved Wilson fermions.} More data at various lattice spacings is required in order to confirm these results. In particular, one needs to check that $\tilde\omega_{1} a $  goes indeed linearly to zero for $a \rightarrow 0$.

The qq+q collaboration employed the plaquette gauge action and the 2-flavor unimproved Wilson quark action. Data was generated at $\beta = 5.1$ ($a\approx 0.195$fm) with four different sea quark masses corresponding to the range $ M_{\pi}/M_{\rho}=  0.47 -  0.76$. Partially quenched data has been accumulated with various valence quark masses for one sea quark mass. For the smallest two sea quark masses, however, $m_{\rm Val}$ had to be chosen equal or larger than $m_{\rm Sea}$ in order to avoid problems with exceptional configurations.

\begin{figure}[t!]
\vspace{-10mm}
\begin{center}
\scalebox{1.55}{\includegraphics*[height=5.5cm,width=5.0cm,angle=-90, bb=50 50 554 604]{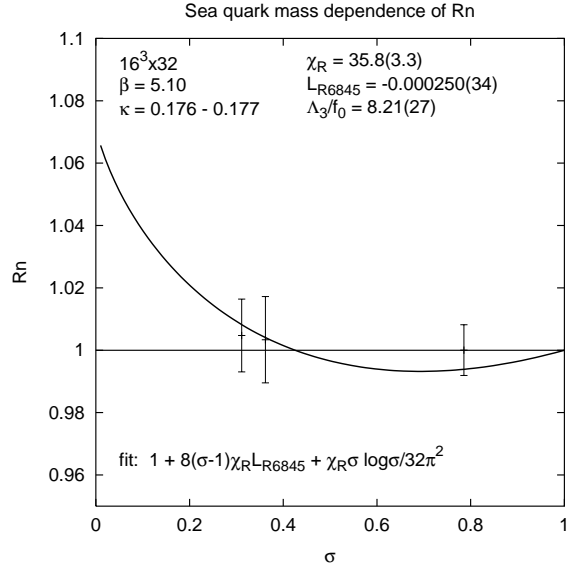}}
\end{center}
\vspace{-10mm}
\caption{Data for $M_{\pi}^{2}/2m_{\rm AWI}$ as a function of $m_{\rm AWI}$, normalized by the values at the heaviest quark mass (see text). From Ref.\ \cite{Farchioni:2004tv}. } 
\label{figqq+q}
\vspace{-6mm}
\end{figure}

Figure \ref{figqq+q} shows the result for the ratio $M_{\pi}^{2}/2m_{\rm AWI}$, normalized by its value at the heaviest quark mass and denoted by $Rn$, as a function of $\sigma= m_{\rm AWI}/m_{\rm AWI,heaviest}$. Note that the normalization by the values at heaviest quark mass disguises the fact that there is a fourth data point at $\sigma=1$. The solid line is the fit result using one-loop continuum \chpt. Similarly to the CP-PACS data the qq+q data can be fitted by continuum \chpt, but the data for the pion mass shows no indication for a curvature due to the chiral logarithms. Even though the data for the pion decay constant shows some curvature \cite{Farchioni:2004tv} more data is needed to corroborate the interpretation in terms of chiral logarithms.

The qq+q collaboration also performed fits using the Wilson \chpt\ expressions including the linear $a$ dependence which were derived in Ref.\ \cite{Rupak:2002sm}. The values for the fit parameters associated with the $a$ contributions turn out to be very small. It was therefore concluded that the lattice artifacts are small.
 
This conclusion seems premature. The lattice spacing $a\approx 0.195$fm and the range of $M_{\pi}/M_{\rho}$ is comparable with the values of the CP-PACS collaboration. As mentioned before, the naive dimensional analysis suggests $m' \approx a^{2}\Lambda_{\rm QCD}^{3}$ for these parameter values. The appropriate fit forms for $M_{\pi}^{2}$ are therefore eqn.\ \pref{1loopAoki} or \pref{1loopAokiResummed}. The fit forms in Ref.\ \cite{Rupak:2002sm} were derived under the assumption  $m' \gg a^{2}\Lambda_{\rm QCD}^{3}$ and are most probably not applicable here. 
Moreover, $m_{\rm AWI}$ was identified with $m'$ (the vector Ward identity mass) in the chiral fit forms. The relation between $m_{\rm AWI}$ and $m'$ is highly non-linear in Wilson \chpt\ with the power counting $m' \approx a^{2}\Lambda_{\rm QCD}^{3}$ and involves chiral logarithms proportional to $am'$ and $a^{2}$.  The simple identification of $m_{\rm AWI}$ with $m'$ is therefore not always justified.
It would be interesting to reanalyze the qq+q data using the proper fit forms for the $m' \approx a^{2}\Lambda_{\rm QCD}^{3}$ regime. This needs to be done before one can draw final conclusions about the size of the lattice artifacts.
%
%

The MILC collaboration has been carrying out 2+1 flavor simulations with staggered fermions, employing a Symanzik improved gauge action and the Asqtad quark action \cite{Aubin:2004fs,BernardLat2004}. Computations have been done for two lattice spacings ($a\approx 0.125$fm and $a\approx 0.09$fm), and fairly small meson masses with $M_{\pi}/M_{\rho}\approx 0.3$ have been reached. Very precise partially quenched  data for the Goldstone boson masses and decay constants have been accumulated with errors of typically $0.1\% - 0.7\%$, and 416 data points are available in total, 208 each for the masses and the decay constants.

Fits of the NLO staggered \chpt\ expressions (the partially quenched analogue of eq.\ \pref{PSMassoneloopSChPT} for 2+1 flavors) to the data give poor results, even if only a subset of 94 data points corresponding to the lightest masses is taken into account. This is not unexpected since the statistical error of the data is much smaller than the estimated uncertainty in the chiral expressions due to neglecting NNLO terms. 

A full NNLO calculation in staggered \chpt\ has not been done yet. Meanwhile, only the analytic NNLO contributions are added to the full NLO chiral fit forms. The total number of unknown parameters in these expressions is 40, which is fairly large. However, only 4 of them are associated with  the non-zero lattice spacing effects. 36 parameters still remain if one sets $a$ to zero in these expressions.

The details of the fits are rather involved, but the bottom line is that good fits are possible with these fit forms (although 176 data points for heavy masses still need to be excluded). One might be tempted to attribute the good fit results to the large number of free parameters. However, good fits are not possible using the fit forms of continuum \chpt, even though the number of free parameters is 36. Similarly, no good fits are possible without the chiral logarithms (38 free parameters). 

The good fits are therefore not a simple consequence of a large number of free parameters. 
The very precise data is able to discriminate between various fit forms, and the results strongly suggest the presence and importance of the taste violating effects of ${\rm O}(a^{2})$. Nevertheless, in order to perform correct fits it seems mandatory to use the complete NNLO expressions including the NNLO chiral logarithms. Recent results in partially quenched continuum \chpt\ at NNLO \cite{Bijnens:2004hk} should help to perform the necessary 2-loop calculations.

\section{Concluding remarks}

The two-step matching procedure to effective field theories (Lattice $\rightarrow$ Symanzik $\rightarrow$ \chpt) has proven to be an appropriate tool for systematically constructing \chpt\ at non-zero lattice spacing. The resulting expressions for physical quantities can differ significantly from the corresponding expressions derived in continuum \chpt, depending on the relative size of the quark mass and the lattice spacing contributions. Expectations from continuum \chpt, in particular with respect to a curvature in lattice data caused by chiral logarithms, might be  misleading. 

Many quantities have been computed to one loop, both in Wilson and in staggered \chpt. Many more calculations remain to be done. Some calculations need to be extended by using a different power counting or by working in the partially quenched approximation. Including the lattice spacing effects in \chpt\  for vector mesons \cite{Jenkins:1995vb} has not been done at all yet. All these calculations need to be done in order to obtain appropriate expressions for the chiral extrapolation at non-zero lattice spacing.  

With the formulation of \chpt\  at non-zero lattice spacing we are able analytically to capture particular lattice artifacts which otherwise would be an uncontrolled uncertainty. The more uncertainties we 
control analytically, the better will be our numerical results for physical quantities.

\vspace{-0.1cm}

\section*{Acknowledgements}

I would like to thank 
S.\ Aoki,
C.\ Aubin,
C.\ Bernard, 
J.\ Bijnens,
C.\ DeTar,
M.\ Golterman,
K.\ Jansen,
I.\ Montvay,
G.\ M\"{u}nster,
Y.\ Namekawa,
G.\ Schierholz,
S.\ Sharpe,
N.\ Shoresh,
R.\ Sommer,
R.\ Van de Water
and
C.\ Urbach
for providing me with material and for helpful discussions.

\end{document}